\documentclass[aps,prb,floatfix,amssymb,amsfonts,amsmath,reprint,superscriptaddress]{revtex4-2}

\usepackage{graphicx}
\usepackage{amsmath}
\usepackage{dcolumn}% Align table columns on decimal point
\usepackage{bm}% bold math
\usepackage{textcomp}
\usepackage{ulem} % strikethrough text

\usepackage{xcolor}

\def\MP{M\textsubscript{2}P}
\def\cm{cm$^{-1}$~}

\begin{document}

\title{Intriguing relaxor ferroelectricity in a polar CT crystal at the neutral-ionic interface}

\author{J. K. H. Fischer} \email{Corresponding author: jfischer@k.u-tokyo.ac.jp}
\affiliation{Department of Advanced Materials Science, University of Tokyo, Tokyo, Japan} 
\affiliation{Experimental Physics V, Center for Electronic Correlations and Magnetism, University of Augsburg, Augsburg, Germany}
\author{G. D’Avino}
\affiliation{Institut N\'eel CNRS \& Grenoble Alpes University, Grenoble, France}
\author{M. Masino}
\affiliation{Dipartimento di Scienze Chimiche, della Vita e della Sostenibilità
Ambientale (S.C.V.S.A.) \& INSTM-UdR Parma, Universit\`a di Parma, 43124 Parma, Italy}
\author{F. Mezzadri}
\affiliation{Dipartimento di Scienze Chimiche, della Vita e della Sostenibilità
Ambientale (S.C.V.S.A.) \& IMEM-CNR, 43124 Parma, Italy}
\author{P. Lunkenheimer}
\affiliation{Experimental Physics V, Center for Electronic Correlations and Magnetism, University of Augsburg, Augsburg, Germany}
\author{Z. G. Soos}
\affiliation{Department of Chemistry, Princeton University, Princeton, NJ 08544, USA}
\author{A. Girlando}
\affiliation{Dipartimento di Scienze Chimiche, della Vita e della Sostenibilità
Ambientale (S.C.V.S.A.) \& INSTM-UdR Parma, Universit\`a di Parma, 43124 Parma, Italy}

\begin{abstract}
We investigated the mixed-stack charge-transfer crystal, N,N’-dimethylphenazine-TCNQ (\MP-TCNQ), which is polar at room temperature and just at the neutral-to-ionic interface (ionicity $\rho \approx  0.5$). We detect the typical dielectric signature of a relaxor ferroelectric and an asymmetric positive-up-negative-down behavior. While relaxor ferroelectricity is usually ascribed to disorder in the crystal, we find no evidence for structural disorder in the investigated crystals. 
To elucidate the origin of \MP-TCNQ's dielectric properties we perform parallel structural and spectroscopic measurements, associated with theoretical modeling and quantum-mechanical calculations.
Our combined effort points to a highly polarizable electronic system that is strongly coupled to lattice vibrations.
The found indications for polarization reversal imply flipping of the bent conformation of the \MP~molecule with an associated energy barrier of a few tens of an eV, broadly consistent with an Arrhenius fit of the dielectric 
relaxation times.
While the polarization is mostly of electronic origin, its possible reversal implies slow collective motions that are affected by solid-state intermolecular interactions. 
\end{abstract}

\maketitle

\section{Introduction}

While spontaneous electrical polarization and ferroelectricity (FE), are well known in organic crystals \cite{Tokura1989}, ~the focus in recent years has ~switched to the~ so-called ``electronic'' FE occurring in charge-transfer (CT) crystals \cite{Nad2006theta,VanDenBrink2008,Horiuchi2008,Tomic2015,Lunkenheimer2015a,Gati2018}. The first examples of electronic FE in organic CT crystals~ came ~from the family of~ tetramethyltetrathiafulvalene (TMTTF) 1:2 salts \cite{Monceau2001,Nad2006,Staresinic2006,Yoshimi2012,Giovannetti2015,Naka2015}. Their FE behavior was initially unexplained, but later suggested to originate from a displacement of the anion chain  accompanied by a shift of the electron holes on the donors \cite{Pouget2012a}.
On the other hand, the most convincing proof of electronic FE,
a $P$-$E$ hysteresis is accompanied by the determination of $P$ orientation with respect to the ion displacement, has so far only been obtained in the ionic phase of tetrathiafulvalene-chloranil (TTF-CA) \cite{Kobayashi2012,Horiuchi2014}, a well-known mixed-stack (ms) CT crystal undergoing the neutral-to-ionic (NI) phase transition at 80 K \cite{Torrance1981,Torrance1981b,Girlando1983,Masino2017}. 
In that system, electronic FE manifested in colossal Born effective charges, whose sign is opposite to that of the valence charge of molecular ions\cite{Kobayashi2012}.
Electronic FE involves the displacement of the molecular $\pi$-electronic clouds, thus implying large values of polarization and fast response to electric fields \cite{Miyamoto2013}. Electronic FE is rare in ms-CT crystals at any temperature, and there is no well documented system at room temperature (RT) \cite{Tayi2012,DAvino2017}. The conditions for electronic FE, i.e., polar crystals close to the NI interface, are indeed difficult to attain.  In the quest for RT electronic FE, we decided to re-investigate an old ms-CT crystal \cite{Soos1977,Soos1978,Nothe1978}, N,N’-dimethylphenazine-TCNQ (\MP-TCNQ),  polar at RT and  just at the NI interface (ionicity $\rho \approx  0.5$) \cite{Meneghetti1985}. While the measurement of a ferroelectric $P$-$E$ hysteresis was not achieved, we detected the typical dielectric signature of a relaxor ferroelectric, already above RT.
Relaxor ferroelectrics are piezoelectric materials characterized by a peak in the $T$-dependent dielectric constant with a pronounced dependence on the frequency of the applied electric field. So far, a precise modeling of relaxor FE is missing \cite{Bokov2006,Cowley2011}. 
The origin of the phenomenon is usually ascribed to the formation of cluster-like, short-range ferroelectric order (polar nanodomains). This may be caused by some form of disorder in the crystal, sometimes artificially introduced \cite{Horiuchi2000}, but there are also examples of nominally well-ordered materials exhibiting relaxor FE, including several CT salts \cite{Matsui2003, Abdel-Jawad2010, Iguchi2013, Lunkenheimer2015}.
In addition to the somewhat unexpected \MP-TCNQ relaxor properties, we also observe an unusual asymmetric positive-up-negative-down (PUND) behavior, pointing to polarization switching.
We investigate the origin of the intriguing dielectric properties of \MP-TCNQ by associating its structural and spectroscopic characterization to the
dielectric measurements. First principle calculations and a simple semimpirical
model are
used to analyze the molecular and collective crystal properties, including electrical polarization, whose origin turns out to be mostly
electronic.  
A plausible scenario is offered to explain the relaxor behavior
and the observed unique PUND features.

\section{Methods}

\subsection{X-Ray diffraction}
Single crystal diffraction data were collected
by using a Bruker D8 Venture diffractometer equipped with a Photon CCD area detector.
CuK$_\alpha$ radiation was used in order to gather reliable information on the absolute structure of the crystal. Low temperature was stabilized by using an Oxford Cryosystems cryostream.
Data reduction was carried out by using the SADABS program \cite{Bruker2008}.
SIR2019 was used for structure solution \cite{burla2015}, and refinement was carried out full-matrix by using the Shelxl program \cite{sheldrick2015}.%\rrr{Sheldrick, G.;
%Crystal structure refinement with SHELXL, Acta Cryst C, 2015, 71(1), 3-8}
The crystal structure was refined making use of anisotropic thermal parameters for all the atoms
except hydrogen, located in the difference Fourier map then constrained during the refinement.

\subsection{Dielectric measurements}

The dielectric constant and conductivity were determined using a frequency-response analyzer (Novocontrol Alpha-A). For the polarization and PUND measurements a ferroelectric analyzer (aixACCT TF2000) was used. Gold wires were attached to contacts of graphite or gold paint on opposite tips of the needle-like crystals, ensuring an electric-field direction exactly parallel to the long crystal axis ($c$). Sample cooling and heating was achieved by a $^4$He-bath
cryostat (Cryovac) and a nitrogen-gas cryosystem (Novocontrol Quatro).

\subsection{Optical spectroscopy}

Infrared (IR) spectra of the crystals were recorded with a
Bruker IFS-66 Fourier transform spectrometer coupled to an IR microscope Hyperion 1000.
Spectral resolution: 2 \cm. We used a wire-grid polarizer and a Polaroid to polarize
the light in the mid-IR and near-IR regions, respectively,
and a gold mirror as reference in the reflectance measurements.
Due to surface irregularities of the samples, the reflectance
values cannot be considered as absolute. The Raman spectra
were recorded with a Renishaw 1000 Raman spectrometer
equipped with the appropriate edge filter and coupled to a Leica
M microscope. Various lines from a Lexel Kr laser were used for excitation.
Incident and scattered polarization was controlled by a half-wave plate and a thin-film linear polarizer, respectively.
A small liquid nitrogen cryostat (Linkam HFS 91) was used for
temperature-dependent measurements under the IR and Raman microscopes.

\subsection{First principles calculations}

Periodic and molecular (quantum chemistry) all-electron Density Functional theory (DFT) calculations employed the global hybrid PBE0 functional in conjunctions with the 6-31G* Gaussian basis set, unless specified otherwise. 
This choice ensures comparable results between the two approaches. 
Periodic DFT calculations were performed with the CRYSTAL17 package \cite{Dov18}.
Quantum chemistry and hybrid quantum/classical (QM/MM) calculations were run with the ORCA code \cite{Neese11}.
Periodic DFT calculations were performed for the $C m$ crystal structure determined in this study at 130 K. Crystal cell parameters were kept fixed to experimental values.
A $2{\times}2{\times}2$ sampling of the Brillouin zone was found sufficient to converge the properties of interest. 

Brillouin-zone center ($\Gamma$ point) lattice dynamics calculations were performed within the harmonic approximations according to established procedures based on the numerical evaluation of the Hessian matrix \cite{Pascale04}.
Empirical Grimme’s D3 pairwise van der Waals corrections \cite{Grimme16} were employed in phonon calculations and in the preliminary geometry optimizations. 
Analytical Raman intensities were computed with a coupled-perturbed Kohn-Sham scheme \cite{Maschio13a,Maschio13b}.
Raman spectra are shown as sums of Lorentzian peaks (half-width at half-maximum of 1.5 cm$^{-1}$), whose amplitude is given by the squared derivative of the polarizability with respect to the normal mode coordinate.
%[ L.MASCHIO, B.KIRTMAN, M.RERAT, R.ORLANDO and R.DOVESI, JCP 139, 164101 (2013)
% L.MASCHIO, B.KIRTMAN, M.RERAT, R.ORLANDO and R.DOVESI, JCP 139, 164102 (2013) ].

Lattice dynamics calculations in soft molecular crystals are %known to be 
extremely sensitive to computational and numerical parameters.
Tight tolerance criteria have hence been used for the convergence of the self-consistent field process ($10^{-12}$ Ha for total energy) and for the optimization of atomic coordinates ($1.2 \cdot 10^{-4}$ bohr and $3 \cdot 10^{-5}$  Ha/bohr for atomic coordinates and gradients, respectively). 
Very high numerical accuracy was requested for the integration grid (XXL grid) and for the truncation of bielectronic integrals (TOLINTEG  8 8 8 8 16). 

A series of lattice dynamics calculations employing the PBE functional were performed to explicitly check whether using a finer sampling of the Brillouin zone ($4{\times}4{\times}4$), or upgrading the basis set by adding polarization function on hydrogens (6-31G**), or by employing triple-zeta functions (6-311G*), leads to modest variations in the vibrational frequencies. 
Specifically, in the most relevant frequency region below 200 cm$^{-1}$,  these parameters determine variations in vibrational frequencies within 5 cm$^{-1}$. 
A similar matching on vibrational frequencies was found between the PBE and PBE0 functional.

The spontaneous electric polarization was calculated within the Berry phase approach \cite{King19}. 
To that end, we built a reference centrosymmetric structure ($C2/m$ space group) and evaluated the variation of the polarization along the path connecting the centrosymmetric and the experimental structure. 
We checked that no discontinuity of multiples of the polarization quantum occurs along this path.
The Berry phase was evaluated for  Kohn-Sham eigenstates, calculated on a $8{\times}8{\times}8$ mesh of the Brillouin zone, ensuring converged results.

Dimerization reversal (flipping) calculations consist of relaxed energy scans, i.e. geometry optimizations where the dihedral angles passing through the central N atoms of \MP\ were constrained to the desired value.
Molecular and solid state relaxed scans were performed with the ORCA and CRYSTAL code, respectively.
The flipping of a single \MP~molecule in the \MP-TCNQ crystal employed hybrid quantum/classical (QM/MM) calculations.
The central QM sub-system (one \MP\ and two neighboring TCNQ along the stack)
was described at DFT (PBE0/6-31G*) level, accounting for the contribution of the MM environment, whose coordinates were kept frozen. 
QM-MM interactions were modelled with the point atomic charges (from electrostatic potential fitting of the DFT density) and van der Waals parameters taken from the GAFF 2010 force field \cite{GAFF2004}.
%\rrr{Comput Chem 25: 1157–1174, 2004}. 
%QM/MM calculations were also performed with the ORCA code.

\section{Results}

\subsection{Structural analysis}\label{x-ray}
We solved the X-ray crystal structure at RT and 130 K, confirming
earlier results \cite{goldberg1973}, but with a better refinement factor: R1 = 0.0303 and 0.0226 at room $T$ and 130 K. %., vs. the previous value of 4.90.
At RT \MP-TCNQ crystallizes in the monoclinic system with $a$=11.1959(5) \AA{}, $b$=13.5747(6) \AA{}, $c$=6.7860(3) \AA{} and $\beta$ =92.436(1)\textdegree. Extinctions affecting the $hkl$ reflections with $h+k=2n$ indices point out C-centering, so that attempts to solve the structure were carried out in the three possible space groups of the corresponding Laue class. The centrosymmetric $C2/m$ and $C2$ space groups produced meaningless results, while the sole reliable solution was found in the polar space group $Cm$ ($C_s^3$), $Z =2$.
Further details of the X-ray structure are reported in Table S1 of the Supplemental Material \cite{Suppmat}.   

No phase transitions are detected by lowering $T$ down to 130 K and the structure retains its RT features, with a slight contraction of the $b$ and $c$ lattice parameters (0.67 and 1.46\% respectively), and a noteworthy increase of $a$ of about 0.3\%. 

No hints of structural disorder are detected, as small, regular, thermal ellipsoids are observed at both temperatures. On the other hand, crystals
with polar point group symmetry usually present inversion twinning, that
we analyzed first by the Shelxl TWIN option.
% was used to account for the inversion twin usually present in crystals with polar point group symmetry \cite{Flack83}.
Both at RT and at 130 K the resulting Flack parameter \cite{Flack83} turns out to be far from the 0 and 1 values expected for untwinned structures. However, the estimates are affected by large uncertainties, likely related to the slight centrosymmetry breaking (i.e. deviation from planar geometry). A more refined symmetry analysis making use of Parsons' method \cite{Parsons13} is less affected by uncertainty, and yields values of 0.34(7)and 0.59(6) for the RT and 130K data collections, respectively. Therefore all the available elements  point to the presence of inversion twinning, although the scale of the twinned domains cannot
be established.

\begin{figure}[htbp]
	\centering
	\includegraphics[width=\linewidth]{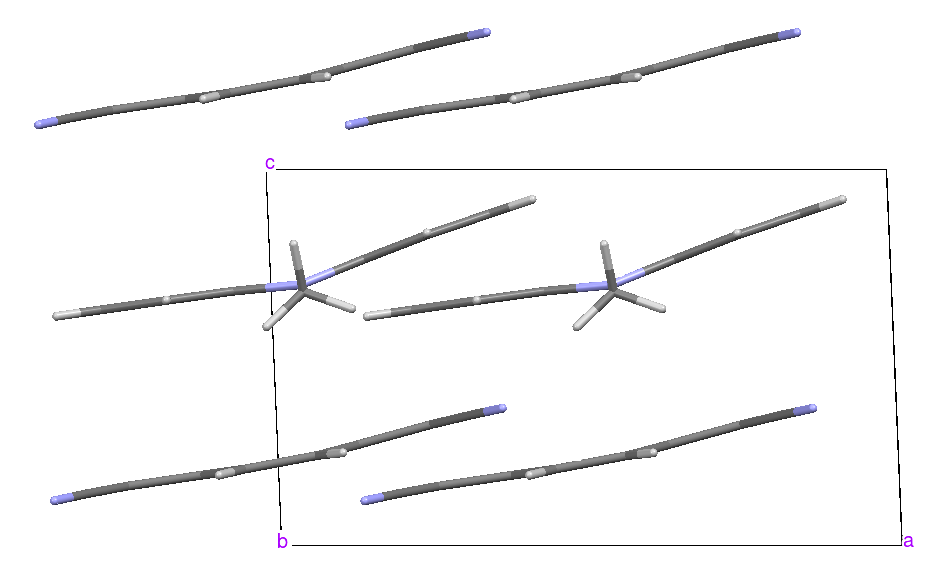}
	\caption{Structure of \MP-TCNQ viewed along the $b$ crystal axis.}
%	Overimposed we show a cartoon view of the intramolecular and intermolecular
%electrostatic dipole moments (dashed fuchsia and green arrows, respectively)
%and the resulting contributions to the cell polarization (block arrows).}
	\label{fig:m2ptqRTstructure}
\end{figure}

Fig.~\ref{fig:m2ptqRTstructure} depicts a projection of the structure
viewed along the $b$ crystal axis. \MP\ appears significantly folded along
the N-N line with a dihedral angle of about 167\textdegree, while
TCNQ is also slightly bent. The dihedral angle of both molecules increases by about one degree upon going to low $T$. The two molecules alternate along the $c$ axis forming a mixed stack where $\pi$-$\pi$ and $\pi$-H interactions seem to dominate. Weak hydrogen bonds affect the inter stack packing, involving the cyanide and methyl groups of TCNQ and \MP\ respectively, while at low $T$ a slightly stronger network of inter-stack interactions sets in, also involving the TCNQ aromatic H atoms (see Fig.~S1) \cite{Suppmat}. 
The DA stacks are dimerized as obviously pointed out by the alternating distances between the molecular centroids $d_1 = 3.341$, $d_2 = 3.513$ \AA~ at 300 K and $d_1 = 3.305$, $d_2 = 3.473$ \AA~ at 130 K. 
The two stacks are dimerized in-phase, i.e., the system is polar, in agreement with the $Cm$ point group symmetry of the crystal, allowing in principle the presence of a spontaneous electric polarization within the $ac$ plane.

The TCNQ distances can be used to estimate the degree of CT or ionicity $\rho$ \cite{hu16}, which turns out to be 0.44. This is less than the $\rho \sim 0.5$ deduced
from IR spectra \cite{Meneghetti1985}, but one has to keep in mind that
the estimates by bond distances are not particularly accurate, especially for $\rho$ appreciably different from zero, and/or when the TCNQ is slightly distorted, like in this structure. In any case X-ray diffraction demonstrates that in going to low $T$ the ionicity does not change.

\subsection{Dielectric measurements}\label{dielectric}

Given its crystal structure, \MP-TCNQ is a good candidate for FE.
To detect polar behavior, dielectric spectroscopy, 
polarization and PUND measurements in a wide frequency and temperature range
were performed, with the electric field aligned exactly parallel to the $c$ stack axis, so that contributions along other directions are not detected.

\begin{figure}[ht]
	\centering
	\includegraphics[width=1.0\linewidth]{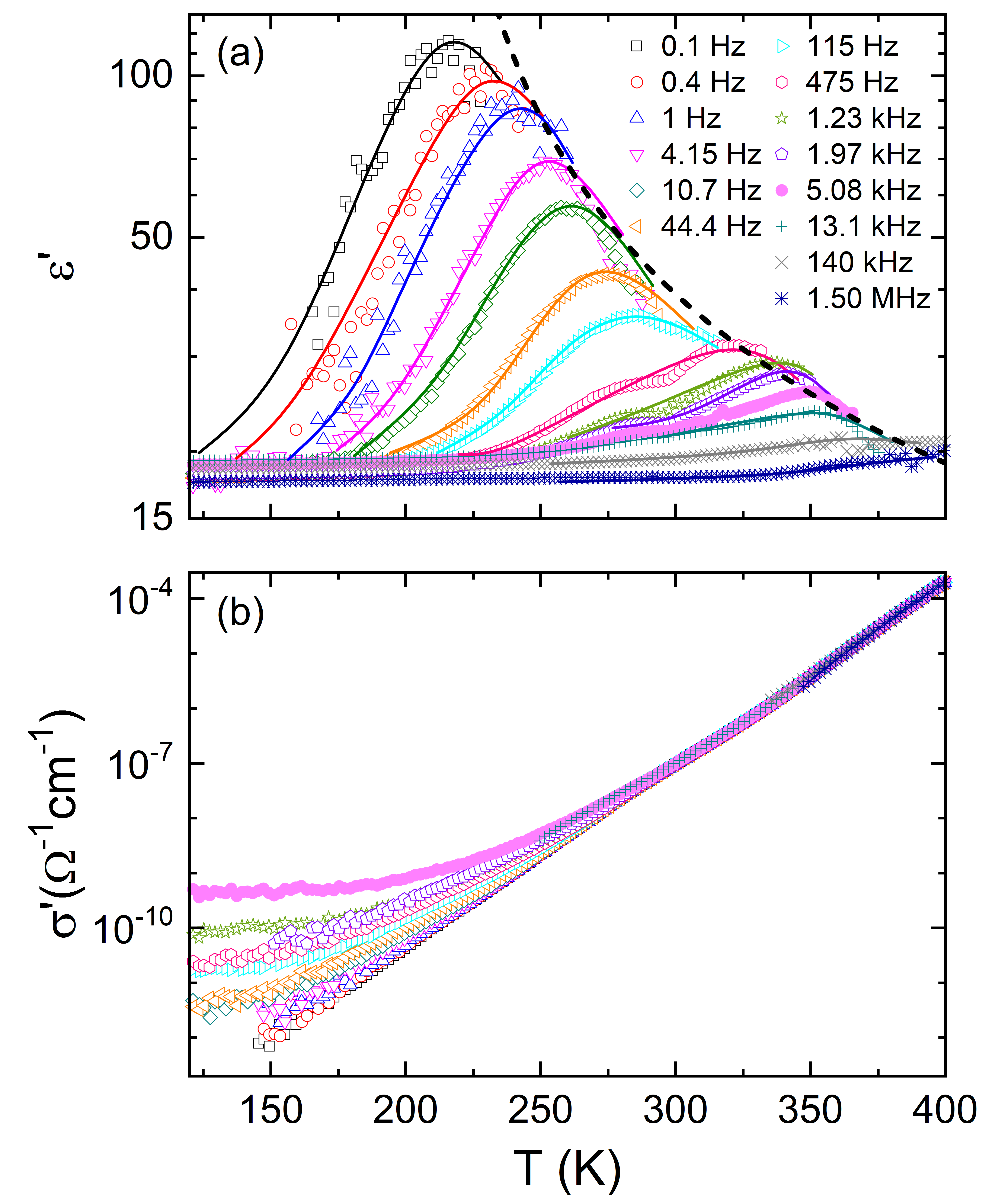}
	\caption{(a) Temperature dependence of the real part of the dielectric constant $\varepsilon'(T)$. The solid lines are guides to the eyes, while the dashed line indicates Curie-Weiss behavior of the right flanks of the peaks, representing the static dielectric constant. (b) Temperature dependence of the conductivity $\sigma'(T)$.}
	\label{fig:m2p-tcnqeps1sigma}
	\end{figure}
In Fig.~\ref{fig:m2p-tcnqeps1sigma}~(a) the temperature-dependent real part of the dielectric constant
$\varepsilon'(T)$ between 0.1 Hz and 1.50 MHz is shown. Large peaks in the permittivity are observed, similar to the peaks at the ferroelectric transition in the related material \MP-Dimethyl-TCNQ (\MP-DMeTCNQ) \cite{Horiuchi1999}. The peaks decrease in amplitude and
shift to higher temperatures with increasing frequency, which constitutes the typical relaxor FE behavior \cite{Cross1987,Samara2003}. In the 0.1 Hz curve the dielectric constant at the peak temperature $T_p$ = 219 K is slightly above 100. At frequencies in the kHz range, the peaks become less and less pronounced, being barely visible in the 13.1 kHz curve, at
$T_p$ = 357 K, with $\varepsilon'(T_p)\approx$ 20.

 To check if the observed relaxor-like behavior of $\varepsilon'(T)$ is intrinsic and not a contact-related artifact \cite{Lunkenheimer2010}, several samples were investigated. Some of the samples were measured with gold paint contacts (flake sizes $\lesssim 10 \mu$m), others with carbon paste (average flake sizes $\simeq 1 \mu$m) (see Fig.~S2 and S3 of the Supplemental Material \cite{Suppmat}). It is indeed well known that the grain size of the metal particles in the paste plays a large role for the formation of Schottky diodes. Therefore, the different contact materials used here, and especially their differing particle sizes, would lead to different results if the relaxor behavior were extrinsic. Additionally, the samples also feature different area-to-thickness ratios, which would lead to marked differences in the dielectric response
of space-charge effects, but not in the intrinsic behavior of the system \cite{Lunkenheimer2010}. All investigated samples do in fact exhibit very similar relaxor-like behavior, thus indicating that the origin of relaxor ferroelectricity in \MP-TCNQ is intrinsic.

 Due to the  needle-like geometry, the electrode area and thus the measured capacitance are very small leading to a large uncertainty of the absolute values of $\varepsilon'$. By comparing different measurements, we obtain a rough estimate for the value of the high frequency dielectric constant $\varepsilon_\infty$ 10,  or somewhat above. Such a value
 is indicative of a relatively high polarizability, as we shall discuss in more detail in Section \ref{dimermodel}.
Note, that at about 300 K a small anomaly of unknown origin is observed at several frequencies, e.g., in the 475 Hz curve, it was also found in the other samples. Finally, the dashed line in Fig.\ref{fig:m2p-tcnqeps1sigma}~(a) is a Curie-Weiss fit to the right flanks of the relaxor peaks, representing the static dielectric constant, with a Curie-Weiss temperature of $T_{\mathrm{CW}} \approx$ 206 K, which provides an estimate of the quasi-static freezing temperature.

\begin{figure}[t]
	\centering
	\includegraphics[width=\linewidth]{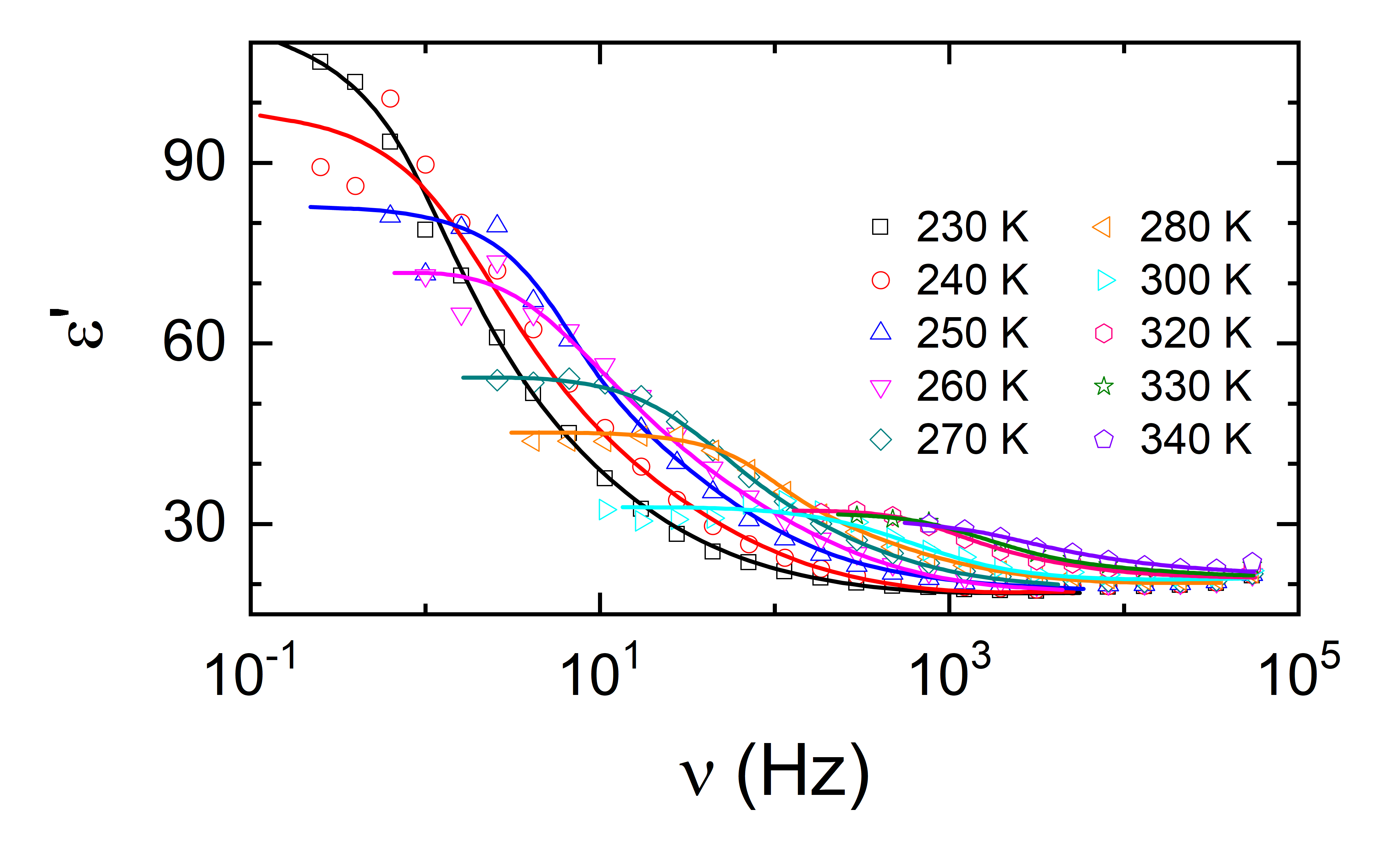}
	\caption{Frequency-dependent plot of the dielectric constant $\varepsilon'(\nu)$ of \MP-TCNQ at various temperatures, revealing  a  step-like  decrease which  shifts  to  lower  frequencies with decreasing temperature. Lines are guides to the eyes.}
	\label{fig:m2p-tcnq-a2f-paper-e1}
\end{figure}

Fig.~\ref{fig:m2p-tcnqeps1sigma}~(b) shows \MP-TCNQ $T$-dependent conductivity $\sigma'(T)$. At 400 K $\sigma' \approx 2 \cdot 10^{-4}~ \Omega^{-1}$cm$^{-1}$ is for all frequencies. With decreasing temperature, the conductivity decreases to about $10^{-7} ~\Omega^{-1}$cm$^{-1}$ at room temperature. Below around 200 K the conductivity becomes slightly frequency dependent pointing to hopping charge transport \cite{Long1982,Elliott1987} whose further investigation is outside of the scope of the present work. At 150 K the conductivity lies between $10^{-9}~\Omega^{-1}$cm$^{-1}$ and $10^{-12}~\Omega^{-1}$cm$^{-1}$. The lowest frequency (0.1 Hz) value can be considered to correspond to the dc conductivity of \MP-TCNQ,
and indeed the value at 150 K, $7 \cdot 10^{-12} ~\Omega^{-1}$cm$^{-1}$, is consistent with the value reported for a compacted polycrystalline sample, $10^{-11} ~\Omega^{-1}$cm$^{-1}$ \cite{Fujita1980}.

A frequency-dependent plot of the dielectric constant $\varepsilon'(\nu)$ is shown in Fig.~\ref{fig:m2p-tcnq-a2f-paper-e1}. The spectra reveal a step-like decrease of $\varepsilon'(\nu)$ which shifts to lower frequencies with decreasing temperature. This evidences the slowing down of relaxational dynamics with decreasing temperature. Similar to the peaks in $\varepsilon'(T)$, the height of curves in $\varepsilon'(\nu)$ decrease with increasing temperature, typical for relaxor ferroelectrics \cite{Cross1987,Samara2003}. In the 230 K curve the highest value of $\varepsilon'(\nu)$ is 110, decreasing to about 30 in the 340 K curve.
\begin{figure}[ht]
	\centering
	\includegraphics[width=0.8\linewidth]{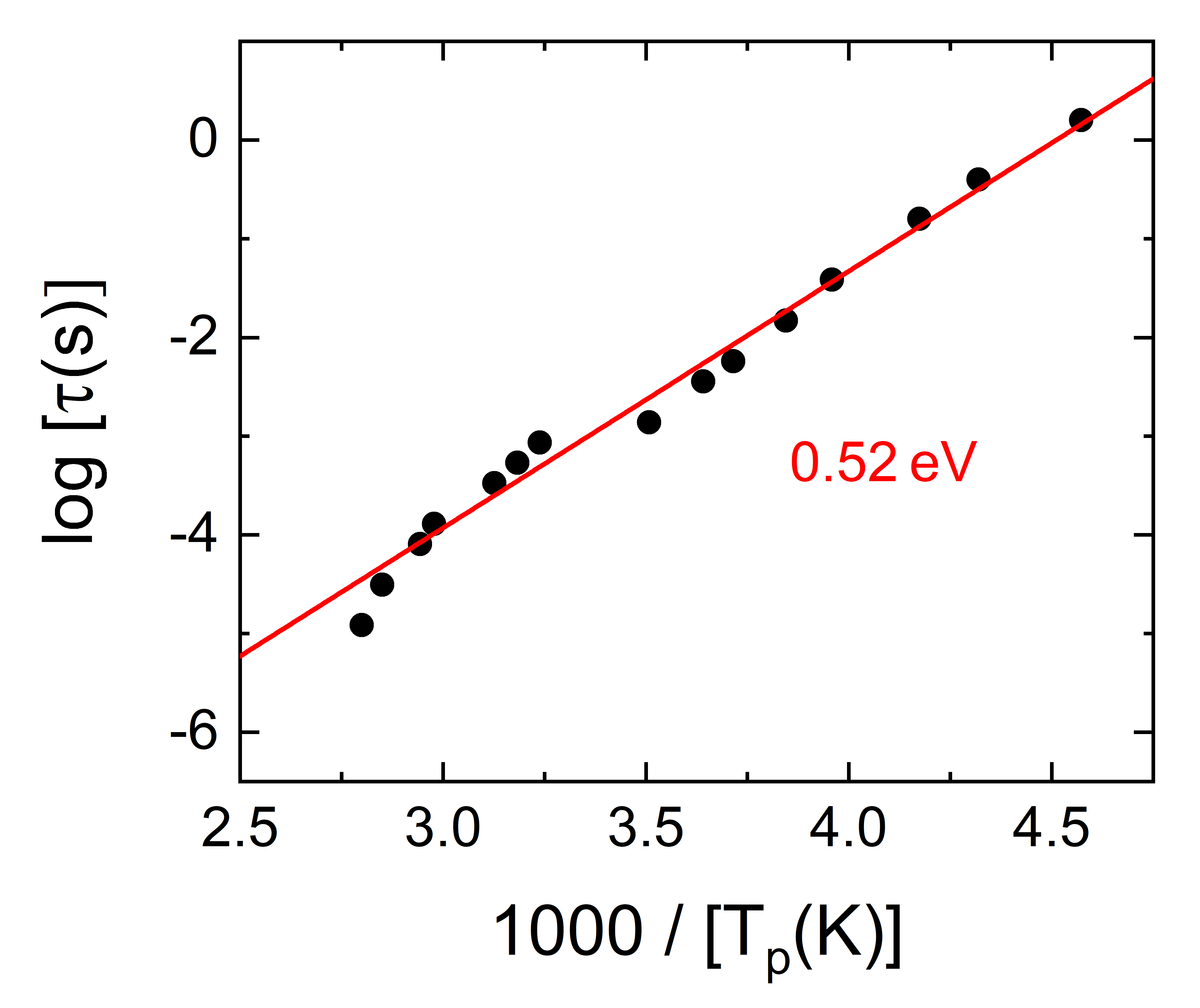}
	\caption{Temperature evolution of \MP-TCNQ relaxation time in an Arrhenius plot. The activation energy is $E_a \approx 0.52$ eV and the pre-exponential factor $\tau_0 = 1.9 \cdot 10^{-12}$ s.}
	\label{fig:m2p-tcnq-arrhenius}
\end{figure}

To further analyze the relaxor ferroelectricity, the peak temperatures 
in $\varepsilon'(T)$ were plotted in an Arrhenius representation, Fig.~\ref{fig:m2p-tcnq-arrhenius}.  The linear Arrhenius fit (red line) to the data yields an activation energy $E_a \approx 0.52$ eV and a pre-exponential factor $\tau_0 = 1.9 \cdot 10^{-12}$ s.
An independent analysis of the points of inflection in the frequency-dependent plot of the dielectric constant $\varepsilon'(\nu)$ in Fig.~\ref{fig:m2p-tcnq-a2f-paper-e1} yields almost identical results.
Although the temperature evolution of the relaxation time of most relaxor ferroelectrics 
can be described by the Vogel-Fulcher-Tammann law \cite{Vogel1921,Fulcher1925,Tammann1900},
%\begin{equation}
%\frac{1}{\tau} = \frac{1}{\tau_0} \exp \left(-\frac{E_a}{\kappa_B (T_{max}-T_f)}\right),
%\end{equation}
%where $T_{max}$ is the temperature of the maximum $\epsilon'$
%and $T_f$ the freezing temperature, 
some relaxors, e.g. PLZT8/65/35 and SBN75 \cite{Kersten1983}, follow Arrhenius behavior. %\cite{Arrhenius1889}.

%:

%\begin{equation}
%\frac{1}{\tau} = \frac{1}{\tau_0} \exp \left(-\frac{E_a}{\kappa_B T}\right).
%\end{equation}
%
%Relaxor ferroelectricity was also reported in several quasi-1D organic compounds, e.g., in TTF-CA doped with QCl$_3$ [horiuchi00], a compound with mixed stacks of donor and acceptor molecules and interchain ferroelectric coupling.

\begin{figure}[hbtp]
	\centering
	\includegraphics[width=\linewidth]{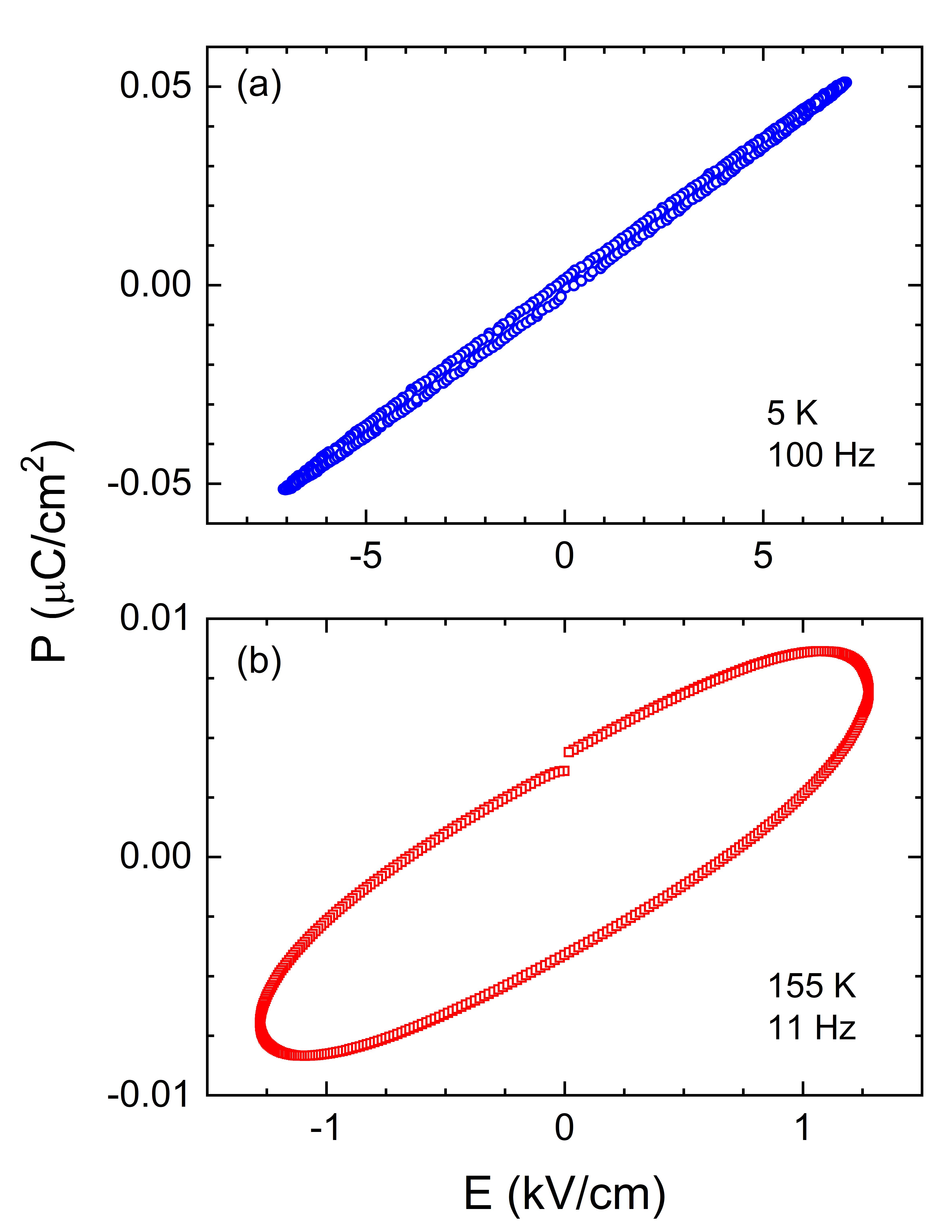}
	\caption{Examples of polarization curves of \MP-TCNQ at (a) 5 K and 100 Hz and (b) at 155 K and 11 Hz.}
	\label{fig:m2ptcnq-pve}
\end{figure}

%\vskip 0.4cm
Polarization measurements performed on several samples of \MP-TCNQ
did not yield a ferroelectric response. The voltage was varied from 20 V to 1000 V (fields up to 50 kV/cm) and the frequency from 0.1 Hz to 1000 Hz at temperatures between 5 K and 270 K. Examples are shown in Fig.~\ref{fig:m2ptcnq-pve}.
At 5 K (upper frame), a nearly linear P-E behavior is observed.
At the higher temperature 
of 155 K, the larger conductivity leads to an ellipsis, typical 
for a sample with some conductivity-related loss. Therefore, we detect well-pronounced polar dynamics, but no ferroelectric hysteresis.  It seems likely that at 5 K, far below the relaxor peaks, the permanent dipoles are essentially frozen-in and cannot be polarized anymore. On the other hand, at higher temperatures the detection of the polarization is hampered by the dominating conductivity contribution. This is a common problem for ferroelectrics that are not perfect insulators.

\begin{figure}
	\centering
	\includegraphics[width=0.8\linewidth]{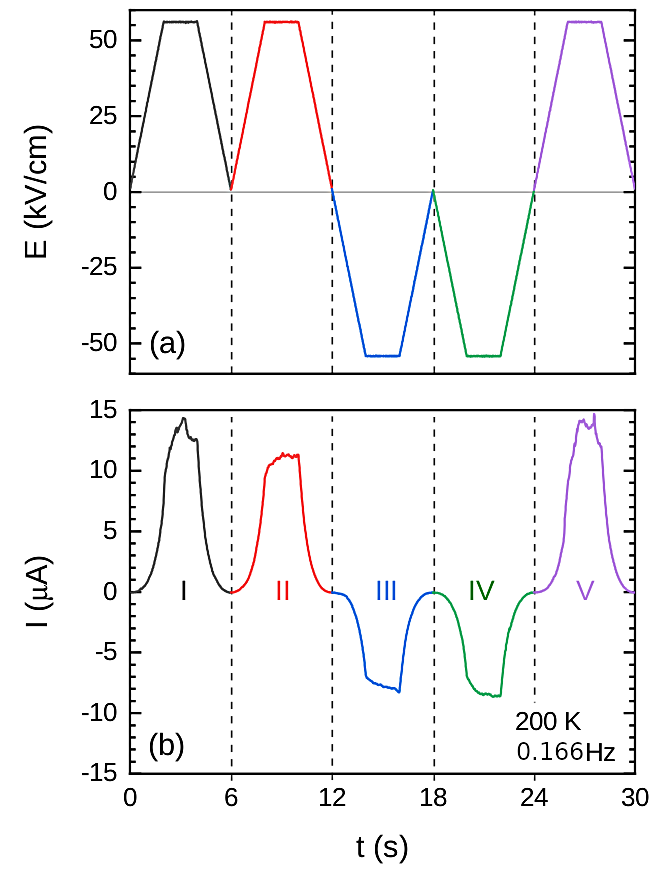}
	\caption{Positive-up-negative-down (PUND) measurement of \MP-TCNQ with (a) applied electric fields of 50 kV/cm resulting in (b) current pulses, of which 1 and 5 are larger than 2. The negative current pulses, 3 and 4, are virtually identical and smaller than the positive ones. Measured at a frequency of 0.166 Hz,  a temperature of 200 K, and with a very short interval between pulses of 1 ms.}
	\label{fig:PUND}
\end{figure}

We also tested polarization dynamics through PUND measurements. Under properly chosen conditions, they exhibit a quite unusual behavior, as shown in Fig.~\ref{fig:PUND}.
The conditions used here are an electric field of 50 kV/cm,
a frequency of 0.166 Hz,  a temperature of 200 K, and a very short interval between pulses of 1 ms (not discernible in Fig.~\ref{fig:PUND}).
These measurement conditions were chosen on the basis of dielectric data shown in Fig.~\ref{fig:m2p-tcnqeps1sigma}.
In order for the conductivity to be small the temperature should be as low as possible. At the same time measuring at a temperature only slightly below the dielectric peak in the spectrum is desirable, since there the ferroelectric correlations of the dipoles can be assumed to be high and the dipoles are sufficiently mobile to be polarizable by an external field. Therefore frequencies below 1 Hz and temperatures below 230 K are the most suitable choice.
In PUND measurements the response of a ferroelectric is expected to feature an additional current contribution for the first of two successive pulses in the same direction (1,3,5), while the second pulse (2,4) is expected to be smaller. This reflects the fact that, after the first pulse, most dipoles are already oriented in field direction and, thus, the second successive pulse does not induce further dipolar motion. In \MP-TCNQ the positive first and fifth current pulses are, as expected, significantly larger than the second one. This is confirmed by calculating the time-integrated current, revealing that the area of peak 2 is about 9\% smaller than those of peak 1 and 5.
However, when applying a negative voltage, both pulses (3 and 4) are virtually identical and the current is about 30\% smaller than in positive direction.

The just described asymmetric PUND behavior of \MP-TCNQ is, to the best of our knowledge, unique. We remark that while we observed similar PUND pattern in another sample, some sample-to-sample difference is present. The observation of the properties depicted in Fig.~\ref{fig:PUND} indeed require carefully chosen measuring conditions, with the employment of large electric fields of at least 50 kV/cm, while higher fields of 60 kV/cm lead to breaking of the samples. 
The breaking could simply be caused by an electric breakdown, but on the other hand it may be an indication that full polarization switching 
is impeded by the molecular geometry of \MP.

We offer the following
plausible but somewhat speculative interpretation: the asymmetry is indicative of preferential one-directional polarization switching, 
since the bending of the \MP ~can not easily be reversed due to 
inter-stack interactions. Forcing the reversal with very strong fields 
involves strong crystallographic rearrangement that breaks the crystal.
Obviously, the dielectric processes in \MP-TCNQ are relatively slow, compared to what would be expected of purely electronic switching \cite{Miyamoto2013}. However, we propose that the rearrangement of the electrons in turn leads to a deformation of the \MP ~ molecules, which slows down the process considerably. This would constitute the reverse of the situation in the (TMTTF)$_2$X salts, where a rigid displacement of the anion chain leads to a shift of the electron holes on the donors \cite{Pouget2012a}.

\subsection{Optical spectra}\label{optical}

\begin{figure}[htbp]
	\centering
	\includegraphics[width=0.8\linewidth]{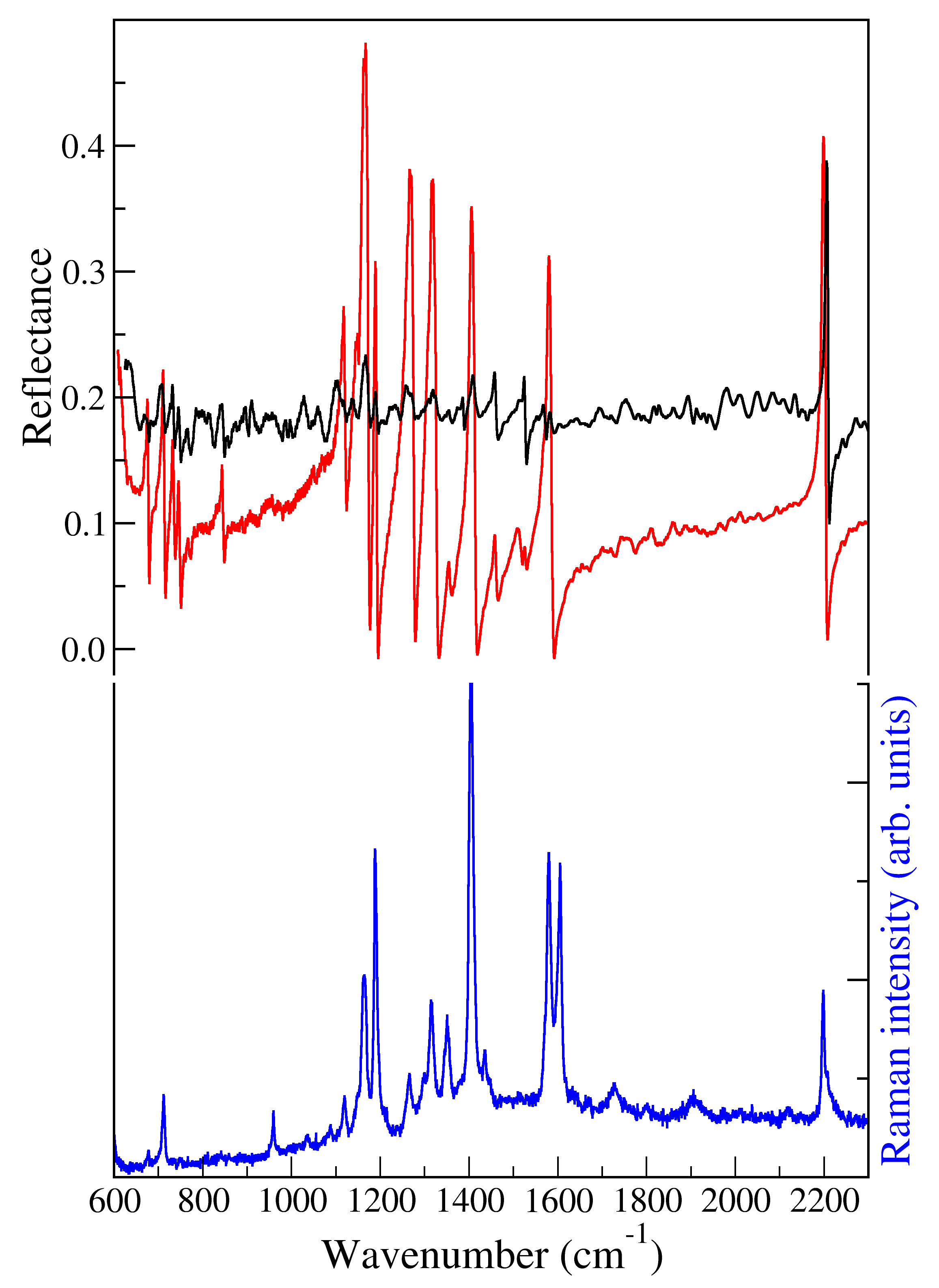}
	\caption{Room temperature polarized IR and Raman spectra of \MP-TCNQ.
		Red line: IR reflectance spectrum, electric vector \textsf{E} $\parallel$ to $c$ axis. Black line: IR spectrum, \textsf{E} $\perp c$. The spectrum is offset for clarity. Blue line: Raman spectrum, ($\perp c$, $\perp c$) polarization, 647 nm 
		%\ag{Matteo check !}
		exciting line.}
	\label{fig:m2ptqrirrt}
\end{figure}

We decided to repeat and extend earlier spectroscopic data \cite{Meneghetti1985}, since optical spectra are a useful complement to the structural and dielectric data.
Examples of the IR and Raman spectra are given in Fig.~\ref{fig:m2ptqrirrt}. 

The IR reflectance spectrum polarized along the stack $c$ axis (red line
in Fig.~\ref{fig:m2p-tcnqeps1sigma}) is dominated by 
very intense transitions at the same frequencies as the main bands observed in the Raman spectrum (blue line). These transitions are due to the electron-molecular vibration ({\it e-mv}) coupling, and confirm
that the stack is dimerized.  On the other hand, the analysis
of the IR spectrum polarized perpendicular to the stack (black line)
allows one to obtain an independent value of the degree of ionicity $\rho$  as $\sim 0.5$, as ascertained by the shift of the TCNQ $b_{1u}\nu_{20}$ mode (C=C antisymmetric stretch). These data fully confirm the previous findings \cite{Meneghetti1985}. An analogous $\rho$ estimate is obtained from the frequency (1404 \cm) of
the Raman active TCNQ $a_g \nu_4$ mode, once the perturbing effect of the {\it e-mv} coupling was taken into account \cite{girlando1985}. The CT transition is also confirmed to occur around 5050 cm$^{-1}$ or 0.63 eV \cite{Meneghetti1985}.

\begin{figure}[htbp]
	\centering
	\includegraphics[width=\linewidth]{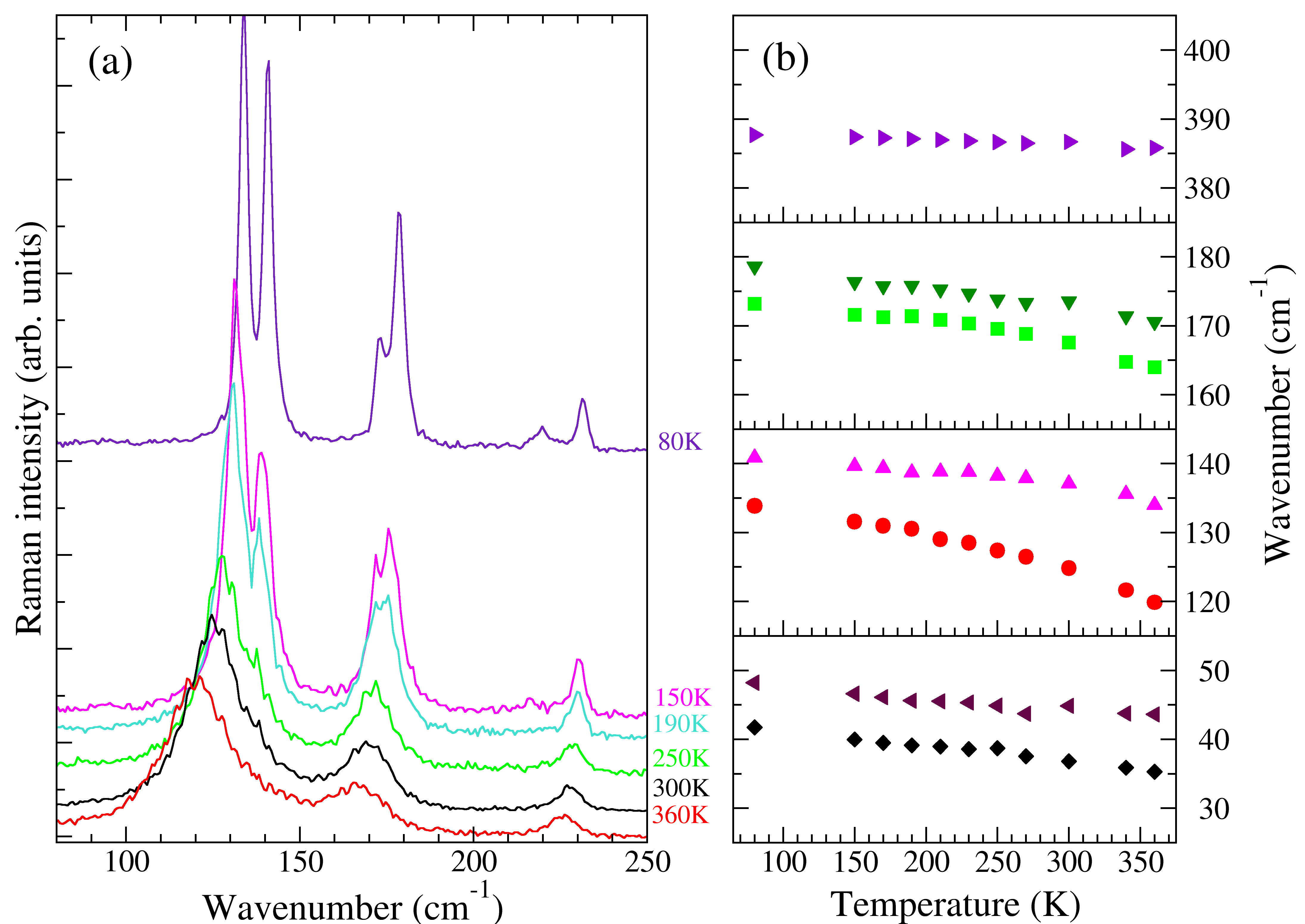}
	\caption{(a) Low-frequency  Raman spectra of \MP-TCNQ, $(c,c$) polarization,  as a function of temperature, exciting line: 752 nm and (b) Temperature evolution of the frequencies of selected bands.}
	\label{fig:Raman_lf_T}
\end{figure}

\begin{figure}[htbp]
	\centering
	\includegraphics[width=0.7\linewidth]{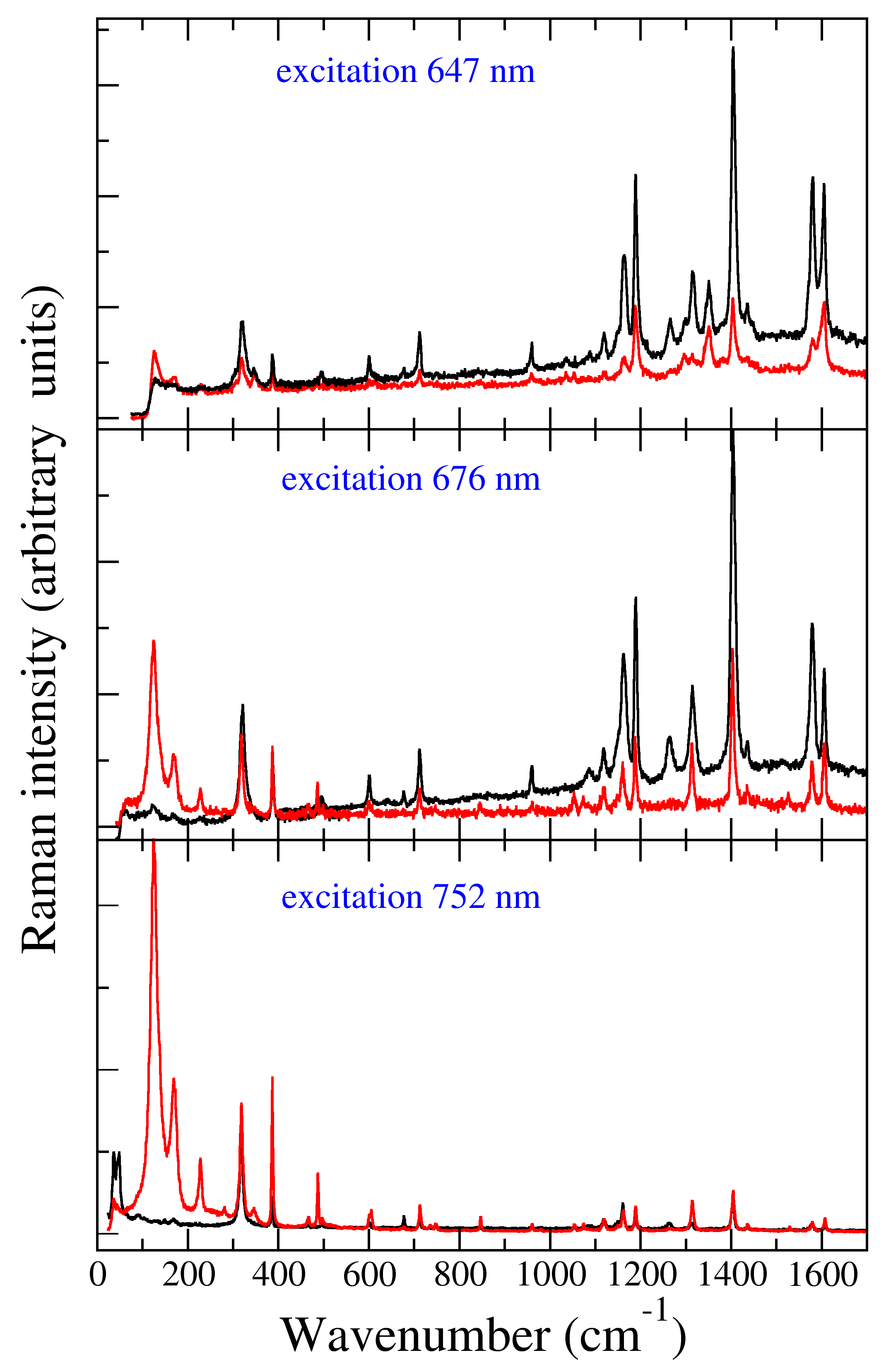}
	\caption{\MP-TCNQ polarized Raman spectra recorded with different exciting lines, $T$ = 300 K. Red line: ($c,c$) polarization; Black line: ($\perp c, \perp c$) polarization.}
	\label{fig:rrsa} 
\end{figure}

We also recorded the spectra as a function of $T$, from 430 to 80 K.
The $T$ evolution of the spectra in the region of the molecular vibrations
shown in Fig.~S4 and S5 \cite{Suppmat} 
demonstrate that there are no phase transitions and that the ionicity or the
extent of dimerization do not change appreciably, as also found in the
X-ray analysis.
New relevant information is instead obtained from the Raman spectra in the
lattice (intermolecular) phonon region (30-200 \cm) shown in Figs. \ref{fig:Raman_lf_T} and \ref{fig:rrsa}.  

Fig.~\ref{fig:Raman_lf_T}~(a) reports the $T$ evolution of \MP-TCNQ low-frequency Raman spectra
with polarization ($c,c$), i.e., incident and scattered radiation with the
electric vector parallel to the stack. 
At room $T$ (black trace in Fig.~\ref{fig:Raman_lf_T}) the spectra
are dominated by two broad bands around 125 and 170 \cm. By increasing the
temperature the bands soften and become broader (bandwidth $\approx$ 35 \cm at 430 K).
By lowering $T$, they narrow considerably and become clearly separated into two bands between
200 and 150 K. To follow the temperature dependence of the frequencies of these two pairs
of bands, we performed a spectral deconvolution in terms of Voigt profiles,
starting from the lowest temperature where all the bands are clearly resolved.
Examples of the deconvolution are reported in Fig.~S6 of the Supplemental Material \cite{Suppmat}.
A stable fitting is found up to 360 K, beyond which the bands become too broad to
give confidence to the result. The central panels of Fig.~\ref{fig:Raman_lf_T}~(b) shows
the temperature evolution of the frequencies of the four bands, evidencing a considerable frequency softening (15-20 \cm) by increasing $T$ in the explored temperature range, a fact pointing to a strong degree of anharmonicity for the associated phonons. The softening
of the other low-frequency phonons is indeed less pronounced (top and bottom
right panels of the Figure) and corresponds to what is normally expected due
to thermal expansion.

The anharmonicity of the four phonon modes associated with the two pairs
of bands around 125 and 170 \cm~ is likely due to a strong electron-phonon coupling \cite{Davino11}.
This idea is confirmed by the collection of spectra with different excitation
lines, shown in Fig.~\ref{fig:rrsa}. It is seen that by shifting the exciting
line towards 
%the red,
longer wavelengths, namely, by going 
%in pre-resonance with 
closer to the CT
transition, the  intensity of the 125 and 170 \cm~ groups of bands
is strongly enhanced with respect to that of the other ones, also
in the different polarization. This resonance intensity enhancement is known to be a consequence
of the electron-phonon coupling \cite{pedron1995}.

\begin{figure}[htbp]
	\centering
\includegraphics[width=0.8\linewidth]{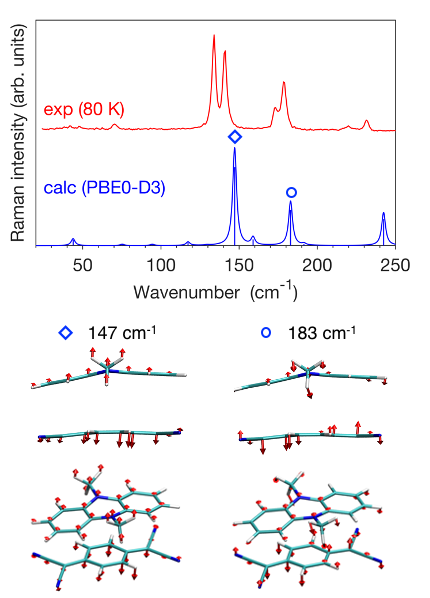} 
	\caption{Experimental (752 nm excitation) and calculated low-frequency Raman spectrum of \MP-TCNQ,
	($cc$) polarization. The eigenvectors of the two most intense Raman bands,
	corresponding to the prominent experimentally observed bands, are shown at the bottom.}
	\label{fig:m2p-tqramancalcexp}
\end{figure}

To properly interpret the lattice phonon spectrum and to gain additional
insight into the electron-phonon coupling mechanism,
we performed DFT calculations of \MP-TCNQ
inter-molecular phonons. The experimental and calculated
spectrum are compared in Fig.~\ref{fig:m2p-tqramancalcexp}.   

The agreement between experiment and calculation can be considered satisfactory, having in mind
%considering
that the experiment is done in pre-resonance with the CT transition, whereas the calculated
Raman intensities are for off-resonance spectra. It is natural
to associate the pair of bands calculated at 147-159 \cm
and 183-192 \cm with the experimental pairs 134-141 \cm and 173-179 \cm
(at 80 K). The eigenvectors of the two pairs are mixed, given the proximity
of their frequency, but in any case the lowest frequency pair is mainly associated
with the dimerization mode (relative displacement of the two sublattices along the stack direction), and the highest frequency one with the ``butterfly''
motion of \MP, as illustrated in the lower part of Fig.~\ref{fig:m2p-tqramancalcexp}. 
Both these motions are involved in a hypothetical, high temperature phase transition towards a paraelectric phase.
As such, these modes are expected to be strongly coupled 
to the electronic CT system, determining the strong anharmonicity that was detected experimentally.

\subsection{First principles calculations}\label{calc}

DFT calculations were performed in order to clarify the mechanism underlying the electrical polarization and its possible switching.
The calculations describe \MP-TCNQ as a band insulator with direct bandgap at the $\Gamma$ point. 
The specific value of the gap markedly depends on the functional, varying between 0.53 eV for PBE (generalized gradient approximation), 0.83 eV for HSE06 (range-separated hybrid), and 1.32 eV for PBE0 (global hybrid).

\begin{figure}[htbp]
\centering
\includegraphics[width=0.8\linewidth]{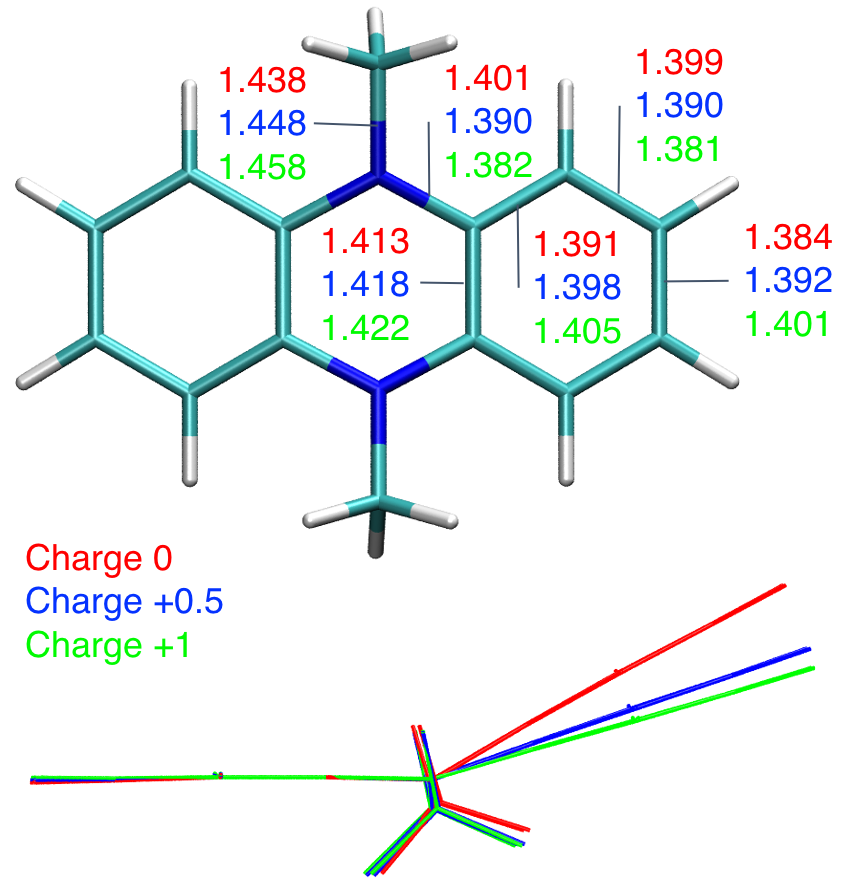}
\caption{Comparison of the equilibrium geometries of an isolated (gas-phase) \MP~neutral molecule (red), cation (green) and for a fractional charge +0.5 (blue), similar to \MP \ in the \MP-TCNQ crystal.
The molecular charge has a modest impact on  bond lengths ($\leq 0.02$ \AA) but it greatly affects the molecular shape, with \MP\ becoming more planar  upon charging. The ``folding-angle'' measures 148\textdegree, 159\textdegree, 163\textdegree~ for 0, +0.5, +1 charge, respectively. 
%Results  from DFT geometry optimization performed at the PBE0/6-31G* level.
Results for +0.5 fractional charge were obtained by considering a symmetric (\MP)$_2^+$ dimer cation, in which the intermolecular distance was constrained to a large value (50 \AA) as to ensure negligible intermolecular interactions.}
\label{fig:m2pchargeflip}
\end{figure}

The stack dimerization of \MP-TCNQ  is intertwined with the folding of the \MP~molecule along the axis passing though the two central N atoms. 
Calculations of the equilibrium geometry in gas phase (Fig.~\ref{fig:m2pchargeflip}) demonstrate that the \MP\ folding depends on the molecular charge, getting more planar upon positively charging. 
The \MP$^+$ cation presents a "folding angle" of 163\textdegree, very similar to the value measured in the \MP-TCNQ crystal where the molecular charge is $\sim$0.5.
Molecular geometries may be strongly affected by intermolecular interactions in the solid state, especially in the presence  of soft degrees of freedom as in the case of \MP.
Interestingly, fully ionic \MP\ can be planar, as seen in the \MP-TCNQF$_4$ CT crystal \cite{Soos1981}.

The considerations above imply that polarization switching would require not only the change in the direction of dimerization, but also the flipping of the \MP\ conformation.
This is specific to this CT crystal with v-shaped molecules that has no counterpart in more common systems (e.g. TTF-CA) featuring molecules that are planar in the gas-phase and only exhibit small deviations from planarity in the crystal. 
This difference is likely to affect the mechanism of polarization reversal, being a possible origin of the relaxor and unique PUND behaviors.

\begin{figure*}[hpt]
	\centering
	\includegraphics[width=0.8\linewidth]{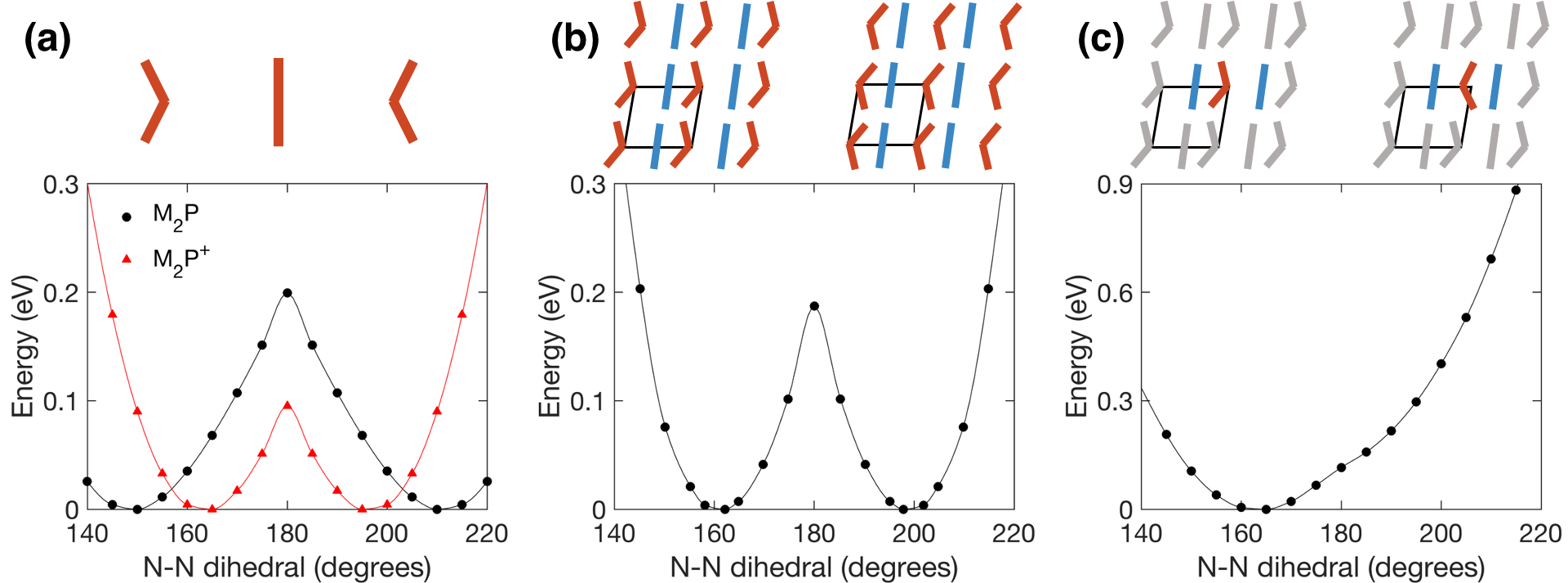}
	\caption{Energy profile to invert the v-like conformation of \MP\ (a) in the gas phase and (b,c) in the crystal. 
		Results from relaxed scans with a constraint on the dihedral angle connecting the two N atoms of \MP. The upper sketches (\MP\ in red, TCNQ in blue) illustrate the flipping procedure.
		(b) Energy scan to simultaneously flip all \MP~molecules in the crystal obtained with periodic DFT calculations.
		(c) Energy scan to flip one \MP~molecules in the crystal from QM/MM calculations. The central \MP~molecule and the two neighboring TCNQ along the stack were relaxed at the DFT level, in the field of the other molecules in the crystal (gray molecules in the sketch) that were kept frozen and described with point charges and Lennard-Jones potential.
		Lines are guides to the eyes.}
	\label{f:flip}
\end{figure*}

We hence calculated the energy profile for flipping the molecule in the gas phase and in the crystal -- see Fig.~\ref{f:flip}.
The results for an isolated \MP, shown in (a), illustrate that the molecular charge affects both the equilibrium geometry (as discussed above, see Fig.~\ref{fig:m2pchargeflip}) and also the energy barrier required to flip the molecule. 
The barrier for \MP\ of 0.2 eV is approximately double that of the cation. 
A barrier of 0.2 eV was also obtained with periodic DFT calculations for the \MP-TCNQ crystal, as shown in Fig.~\ref{f:flip}~(b). 
This corresponds to the energy barrier to invert the polarization in a single-domain macroscopic crystal.
We note that since this barrier is much larger than the room-temperature thermal energy (26 meV), the polarization reversal appears to be energetically impeded.
On the other hand, the barrier is comparable in magnitude with, yet significantly smaller than, the 0.5 eV estimated from the Arrhenius fit of the dielectric data depicted in Fig.~\ref{fig:m2p-tcnq-arrhenius} (Section \ref{dielectric}).

We also considered a third possibility, in which a single \MP~molecule is flipped in the crystal that retains the original polarization -- see Fig.~\ref{f:flip}~(c).
This situation corresponds to the creation of a defect in the otherwise periodic crystal structure, and was modelled with hybrid quantum/classical calculations (QM/MM, see Methods).
In this case, the only stable configuration remaining is that with all \MP~molecules pointing in the same direction, since the second conformation is strongly destabilized by intermolecular interactions in the solid.
Such a destabilization is imputable to dispersion interactions and steric repulsion and not to an electrostatic effect associated with the dipole reversal on the central molecule. Indeed, a very similar energy profile is obtained by neglecting electrostatic interactions with the MM environment.

As stated above, polarization inversion requires both the inversion of the dimerization and the flipping of \MP. 
We may expect that phonons associated with those motions are the most strongly coupled to the electronic CT system, hence the most anharmonic, which we identified experimentally with the two groups of bands around 125 and 170 \cm (RT frequencies). 

The bent shape of \MP\ also has important consequences on the amplitude  of dimerization of the mixed stack.
In the context of lattice models for the electronic structure, the dimerization amplitude is usually defined as $\delta=(t_1-t_2)/(t_1+t_2)$, %$\delta=\frac{t_1-t_2}{t_1+t_2}$, 
where $t_1$ and $t_2$ are the charge transfer integrals between neighboring donor and acceptor molecules along the stack.
The two limiting cases $\delta=0$ and $\delta=1$ correspond to a regular stack of equally spaced molecules and to a crystal of non-overlapping dimers, respectively.
DFT calculations (PBE0/6-31G* level, projective method \cite{Valeev06}) on \MP-TCNQ dimers extracted from the crystal structure yield $t=(t_1+t_2)/2= 493$ meV and $\delta=0.32$. 
It is known that the magnitude of transfer integrals in donor-acceptor complexes strongly depends on the functional employed, and in particular on the fraction of Hartree-Fock exchange $\alpha_{HF}$ \cite{Sini11}.
Within the framework of the one-parameter global hybrid functionals of the PBEh($\alpha_{HF}$) family, a variation of $\alpha_{HF}$ by $\pm0.1$ affects $t$ by $\pm0.12$ eV.
On the other hand, the dimerization amplitude $\delta=0.32$ is weakly functional dependent.
This value for \MP-TCNQ is considerably larger than that calculated for the ionic phase of TTF-CA at the same level of theory, $\delta=0.19$. 
This remarkable difference between the two systems suggests that \MP-TCNQ, owing to its non-planar shape of the donor molecule, is more suited to be described as a collection of weakly overlapping dimers.

We then calculated the electric polarization of the \MP-TCNQ crystal. %, defined with respect to a reference centrosymmetric structure. 
These calculations aim at predicting the $\mathbf{P}$ that one would measure if it were possible to switch the polarization as in a normal ferroelectric phase, such as in TTF-CA \cite{Kobayashi2012}.
The data in  Table~\ref{tab:polarization} reveal that the polarization of \MP-TCNQ has two components along $a$ and $c$ that are comparable in magnitude.
This surprising result reveals another intriguing difference with respect to other mixed stacks with planar molecules (TTF-CA, TTF-QBrCl$_3$, TTF-BA), for which the polarization is essentially directed along the stack axis \cite{Horiuchi2014}.
In order to understand the physical origin of the polarization, Table~\ref{tab:polarization} also reports the ionic polarization (calculated considering charges $\pm0.5e$ at the molecular centroids, as in Ref. \cite{Kobayashi2012}) and the dipolar polarization.
The latter accounts for the contribution of the dipole moments of the individual neutral molecules at the crystal structure geometry, calculated with gas-phase DFT.
Ionic and dipolar contributions are both automatically included in the DFT calculation of the total polarization.

\begin{table}[h]
	\centering
	\caption{Components of \MP-TCNQ electrical polarization $\mathbf{P}$ calculated with periodic DFT and with classical models. 
		DFT results include both electronic (Berry phase) and nuclear contributions. 
		The table reports the components along $a$ and $c$ crystal axes (forming an angle of 92\textdegree). The polarization along $b$ is zero by symmetry.}
	\begin{tabular*}{0.48\textwidth}{@{\extracolsep{\fill}}lcc}    	
		\hline \hline
		Model  & $P_a$ ($\mu$C/cm$^2$) & $P_c$ ($\mu$C/cm$^2$)\\
		\hline 
		PBE0             & -5.88 & -5.16 \\ 
		ionic            & -0.87 & 0.04 \\  % center of charge (Loewdin)  
		dipolar          & -0.02 & 0.40 \\
		\hline \hline
	\end{tabular*}
	%The ionic model considers charges $\pm0.5e$ at the center of charge of molecular ions. 
	%The dipolar model accounts for the contribution of the dipole moments of the individual molecules at the crystal structure geometry, calculated with gas-phase DFT.
	%}
	\label{tab:polarization}
\end{table}{}

The stack-axis component of the polarization, $|P_c|=5.16$~$\mu$C/cm$^2$, is similar to what was measured for the ionic low-$T$ phase of TTF-CA. 
The ionic polarization $P_c$ is two orders of magnitude smaller that the total polarization and points in the opposite direction, marking an important analogy with TTF-CA and TTF-QBrCl$_3$ \cite{Pion}.
These similarities concerning the stack-axis polarization point to a common electronic mechanism of polarization, governed by fluctuations of electronic charges along the stack, rather than by the frozen ionic charges localized at molecular sites in a dimerized lattice.
The dipolar contribution along $c$ is non negligible, allowing us to obtain an estimate of the electronic polarization by subtraction,
$P_c^{(el)}=P_c-P_c^{(ion)} - P_c^{(dip)}=-5.6$~$\mu$C/cm$^2$.

As anticipated, the leading component $P_a$ has no counterpart in traditional mixed-stack crystals.
In this case the ionic and total polarization are parallel, with the former accounting for 15\% of the total one.
The ionic contribution along $a$ is much larger in magnitude than its component parallel to $c$.
This results from a more pronounced displacement of the donor and acceptor sublattices (with respect to a centrosymmetric arrangement) along $a$, that can be also inferred from the visual inspection of the crystal structure in Fig.~\ref{fig:m2ptqRTstructure}.
Along $a$ the dipolar contribution is negligible, leading to an electronic polarization $P_a^{(el)}=-5.0$~$\mu$C/cm$^2$, slightly smaller than $P_c^{(el)}$.

Finally, we calculated the molecular polarizabilities at the PBE0/ma-def2-TZVP level and obtained  $\alpha \approx $~60~\AA$^3$ for either neutral molecules, $\alpha(\mathrm{M_2P}) + \alpha(\mathrm{TCNQ})$, or molecular ions, $\alpha(\mathrm{M_2P^+}) + \alpha(\mathrm{TCNQ^-})$. 
The calculated polarizability of a dimer taken from the crystal structure is $\alpha(\mathrm{M_2P^{+\rho}TCNQ^{-\rho}}) \sim 120$ \AA$^3$ with $\rho = 0.44$ close to the experimental $\rho \sim 0.5$. Since molecular polarizabilities are too small to account for the $10-15$ value of $\varepsilon_\infty$ in Section \ref{dielectric}, we consider polarizability due to the crystalline environment.

\subsection{A simple model for strongly dimerized mixed stacks}\label{dimermodel}

The first principles calculations of the previous Subsection put in evidence the prevailing electronic origin of the polarization of the \MP-TCNQ crystal. 
Electronic FE requires a highly polarizable lattice.
Here we focus on the microscopic origin of the high polarization through a simple semiempirical model.
The reference model for the electronic structure of ms-CT crystals is a Peierls-Hubbard model with staggered site energies for donor (D) and acceptor (A) sites and long-range Coulomb interactions \cite{Davino17}.
The large dimerization of \MP-TCNQ stacks, however, suggests that a first approximation to the crystal described is non-overlapping donor-acceptor (D-A) dimers along the stack 
with 3D Coulomb interactions with all other dimers.
The mean-field treatment of the interactions between dimers leads to the so-called embedded Mulliken dimer model that was introduced in Ref.~\cite{Soos1978} and recently reviewed \cite{Davino17}.
% We focus here on implications of embedding on dielectric properties.
 
The Mulliken dimer model describes an isolated dimer on the basis of neutral $|\mathrm{DA}\rangle$ and ionic $|\mathrm{D^+A^-}\rangle$ electronic states. The Hamiltonian 
in the singlet sector reads:
\begin{equation} \label{eq:Hda}
    H_0=2z_0\hat \rho -\sqrt{2}t \hat\sigma_x 
\end{equation}
where $2z_0=(\mathcal{I}-\mathcal{A}-V)$, $\mathcal{I}$ is the D ionization potential, $\mathcal{A}$ is the A electron affinity, $V$ is the nearest neighbor Coulomb interaction and $t$ is the CT integral. 
$\hat\rho =(1-\hat\sigma_z)/2$ is the ionicity operator, where
$\hat\sigma_z$ and $\hat\sigma_x$ are the Pauli matrices.
Having defined the dipole moment operator as $\hat\mu=ea\hat\rho$, where $e$ is the electron charge and $a = 3.5$ \AA~ is the intermolecular spacing, the polarizability of the isolated dimer can be expressed in terms of the ground state ionicity:
\begin{equation} \label{eq:alpha0_rho}
\alpha_0(\rho)=\frac{2(ea)^2}{\sqrt{2}t}(\rho(1-\rho))^{3/2}.
\end{equation}

The Hamiltonian for an embedded dimer is formally equivalent to Eq.~\ref{eq:Hda} with $z_0$ replaced by $z=z_0-\epsilon_c\rho$, where
\begin{equation}
\epsilon_c=M-V,
\end{equation}
and $M$ is the Madelung energy of the ionic ($\rho = $1) lattice.
The embedded dimer Hamiltonian depends self consistently on $\rho$, which accounts for cooperative inter-dimer interactions in the solid-state.
Positive $\epsilon_c$ values correspond to attractive interactions between dimers, favoring the ionic state.
The model describes the crossover between a neutral and an ionic ground state upon decreasing $z(\rho)$.
The crossover is continuous at $\rho=0.5$ ($2z=\mathcal{I}-\mathcal{A}-V- \epsilon_c$) for 
$\epsilon_c<2\sqrt{2}t$, while a first-order transition with a phase coexistence region is obtained for $\epsilon_c>2\sqrt{2}t$. 
The polarizability of the embedded dimer is
\begin{equation}\label{eq:alpharho}
\alpha(\rho) = %\frac{ea}{2} \frac{d\rho}{dz_0}=
\frac{\alpha_0(\rho)}{1 - [2\epsilon_c\alpha_0(\rho)]/(ea)^2] }.
\end{equation}
This self-consistent analytical
%\gd{I would call it "analytical result/expression" or just "result/expression"}
result was previously obtained for an analogous Hamiltonian that describes the vibrational enhancement of the electric susceptibility of push-pull chromophores \cite{Painelli99}.

Eq. \ref{eq:alpharho} is general and together with Eq. \ref{eq:alpha0_rho}
shows that the polarizability of the embedded dimer is maximum for intermediate
$\rho \sim 1/2$ as in \MP-TCNQ. 
Since $\rho(1-\rho) = 1/4 - (\rho-1/2)^2$,
% \gd{I find this explanation unnecessary, but not harmful}
 we get
\begin{equation}\label{eq:alpha05}
 \alpha(0.5)=  \frac{(ea)^2}{2[2\sqrt{2}t -(M-V)]}, 
\end{equation}
i.e. the polarizability from intermolecular CT degrees of freedom may become very large on approaching the critical point $(M-V) = 2\sqrt{2}t$, where it diverges.

In order to check the prediction of the model, we of course need to estimate the relevant parameters. 
According to the embedded dimer model, the energy of the CT transition for $\rho = 0.5$ is $E_{\mathrm{CT}} = 2\sqrt{2}t$, so that  $ t\simeq 0.2$ eV. 
We followed the method described in Ref.\cite{Delchiaro18} to calculate $\epsilon_c = M-V = 2.6 - 2.2$ eV $ = 0.4$ eV.
We are indeed relatively close to the critical point separating continuous from discontinuous crossover, and from Eq. \ref{eq:alpha05} we get $\alpha \sim 800$ \AA$^3$.

Always within the embedded dimer model, we can also estimate $\alpha$ from the experimental value of $\varepsilon_\infty \approx 10 - 15 $ and
the relation $\varepsilon_\infty = 1+4\pi\alpha/V_{DA}$, where $V_{DA} = 502$ \AA$^3$
is the volume occupied by the a DA dimer. We obtain $\alpha \approx$ 400 to 600 \AA$^3$,
a value consistent with that derived from Eq. \ref{eq:alpha05}.
Therefore, a major contribution to the dielectric response arises from intermolecular CT degrees of freedom, enhanced by solid-state electrostatic interactions.
This simple treatment gives some general clues about the microscopic requirements to achieve highly polarizable electronic systems, and hence promising candidates for electronic FE.

\section{Discussion and conclusion}

The performed dielectric measurements of \MP-TCNQ, a polar mixed-stack charge transfer crystal, provide clear evidence for relaxor FE behavior already above room temperature (Fig.~\ref{fig:m2p-tcnqeps1sigma}).
The analysis of the temperature-dependent relaxation time, derived from the frequency dependence of the dielectric constant,
reveals an Arrhenius behavior with an energy barrier of  $E_a \simeq 0.52$ eV (Fig.~\ref{fig:m2p-tcnq-arrhenius}).
Even with an electric fields up to 50 kV/cm, we did not observe polarization switching, while larger fields lead to the breaking of the crystal. Under carefully chosen measuring conditions, we detect a unique asymmetric PUND behavior (Fig.~\ref{fig:PUND}), where the positive pulses exhibit the behavior expected of a FE,  while the negative pulses are significantly smaller. 

Literature optical data \cite{Meneghetti1985}, fully confirmed by the present ones, indicate that \MP-TCNQ is among the rare crystals at the borderline between neutral and ionic ground state, with ionicity $\rho \sim 0.5$ that does not appreciably change with temperature from 400 to 80 K.
Raman measurements in the low-frequency spectral region put in evidence
the presence of two pairs of anharmonic phonons, around 150 and 180 \cm, likely
coupled to the electronic system. DFT calculations demonstrate that these phonons
involve the dimerization (Peierls) mode, and the butterfly motion of the bent
\MP~molecule. These modes are intertwined through the common interaction with the
CT, and would be involved in a hypothetical transition to a high-temperature paraelectric phase. 

Polarization switching in \MP-TCNQ requires the flipping of the \MP~molecule, which implies an energy barrier of 0.2 eV for the isolated molecule (DFT estimate), which, however, might differ significantly in the solid state as a result of intermolecular interactions.
This marks an important qualitative difference with respect to the prototypical CT ferroelectric TTF-CA, for which the polarization reversal implies only the rigid translation of planar molecules.
First principles calculations reveal that the electrical polarization of \MP-TCNQ is of \textit{quantum electronic} nature, as in TTF-CA \cite{Giovannetti09}.
The magnitude of the stack-axis component (5.2 $\mu$C/cm$^2$) is similar to the one measured for TTF-CA and TTF-QBrCl$_3$ \cite{Kobayashi2012}.
Thus the possible ferroelectricity of \MP-TCNQ is of {\it electronic origin},
but does not constitute purely electronic {\it switching} due to the involvement of the bending \MP~molecule.

\MP-TCNQ offers an intriguing and challenging experimental scenario, especially concerning the understanding of the relaxor behavior.
Relaxor ferroelectricity is often associated with some kind of disorder \cite{Cowley2011}.
In molecular salts disorder was introduced artificially in a controlled way \cite{Horiuchi2000}, 
yet relaxor behaviour was also observed in pristine systems, e.g., $\lambda$-(BEDT-TSF)$_2$FeCl$_4$ \cite{Matsui2003}, $\kappa$-CN \cite{Abdel-Jawad2010}, and $\alpha$-(ET)$_2$I$_3$ \cite{Lunkenheimer2015}.
The absence of structural disorder in our samples does not exclude the possibility of disorder on a smaller scale, like charge defects or nanodomains with opposite dipolar orientation, as suggested by the intermediate value of the Flack parameter in the X-ray analysis. Another possible origin of relaxor behavior is the bent geometry of \MP~molecules, as shown in Fig.~\ref{fig:m2pchargeflip}.

Based on the available data, it seems plausible that the domain-wall motion under the action of an electric field is extremely slow in \MP-TCNQ as compared to mixed-stack CT crystals of planar molecules, such as TTF-CA. 
The latter exhibits clean hysteresis loops \cite{Kobayashi2012}, and full sample poling at fields of 0.95 kV/cm, after which the sizeable contribution of domain walls (soliton) motion to the dielectric constant in the ionic phase is suppressed \cite{Kagawa10}. 
The bent shape of \MP\ with its two stable conformations separated by a large energy barrier, which is probably enhanced by intermolecular interaction at the domain boundary, is likely to be the origin of the singular behavior of this material.
We therefore expect that the relaxation dynamics associated with domain-wall motions depend on the specific features of the domain boundary and its interplay with structural defects, which may provide a rationale for the broad spectrum of characteristic timescales observed by dielectric spectroscopy.
As a rare example of a ms-CT salt that is relaxor ferroelectric above room temperature, the details of \MP-TCNQ's electronic ferroelectricity deserve further attention in future studies.

\section*{Acknowledgments}
We thank Prof. Anna Painelli for enlightening discussions.
JKHF and PL acknowledge funding from the Deutsche Forschungsgemeinschaft (DFG) via the Transregional Collaborative Research Center TRR80 (Augsburg, Munich). JKHF was supported by JSPS [Postdoctoral Fellowships for Research in Japan (Standard)] as International Research Fellow. In Parma the work has benefited from the equipment and support of the COMP-HUB Initiative, funded by the ``Departments of Excellence'' program of the Italian Ministry for Education, University and Research (MIUR, 2018-2022).

\bibliography{m2ptq_8}% Produces the bibliography via BibTeX.

\end{document}